%% file: main.tex
\documentclass[sigconf]{acmart}

\usepackage{tabularx} 
\usepackage{multirow}

\acmDOI{}

\acmISBN{}

\acmConference[RAISE'18]{6th International Workshop on Realizing Artificial Intelligence Synergies in Software Engineering}{May 2018}{Gothenburg, Sweden}
\acmYear{2018}
\copyrightyear{2018}
\acmArticle{} 
\acmPrice{}  

\begin{document}

\title{Ways of Applying Artificial Intelligence in Software Engineering}

\author{Robert Feldt, Francisco G.\ de Oliveira Neto, and Richard Torkar}
\affiliation{%
  \institution{Chalmers and the University of Gothenburg}
  \city{Gothenburg}
  \country{Sweden}
}
\email{[robert.feldt|gomesf|torkarr]@chalmers.se}



\begin{abstract}
As Artificial Intelligence (AI) techniques have become more powerful and easier to use they are increasingly deployed as key components of modern software systems. While this enables new functionality and often allows better adaptation to user needs it also creates additional problems for software engineers and exposes companies to new risks. Some work has been done to better understand the interaction between Software Engineering and AI but we lack methods to classify ways of applying AI in software systems and to analyse and understand the risks this poses. Only by doing so can we devise tools and solutions to help mitigate them. This paper presents the AI in SE Application Levels (AI-SEAL) taxonomy that categorises applications according to their \textit{point} of AI application, the type of AI \textit{technology} used and the \textit{automation level} allowed. We show the usefulness of this taxonomy by classifying 15 papers from previous editions of the RAISE workshop. Results show that the taxonomy allows classification of distinct AI applications and provides insights concerning the risks associated with them. We argue that this will be important for companies in deciding how to apply AI in their software applications and to create strategies for its use.
\end{abstract}

%
%


\maketitle

\input{./files/introduction}
\input{./files/background}
\input{./files/ai_seal_taxonomy}

\input{./files/evaluation}
\input{./files/conclusions}

  

\bibliographystyle{ACM-Reference-Format}
\bibliography{sample-bibliography}

\end{document}

%% file: files/introduction.tex
\section{Introduction}
\label{s:introduction}

Artificial Intelligence (AI) has shown a lot of promise in the last decades but with the recent resurgence of interest and improved results on real-world tasks the field is undergoing explosive growth. Many of the improved results have come from larger and more complex neural networks, stacked many layers deep (for so called Deep Learning), but much of progress can also be attributed to larger data sets and large-scale learning\slash training on GPUs~\cite{schmidhuber2015deep}. But the renewed interest and increasing amount of resources has also lead to breakthroughs in related AI technologies, e.g. Bayesian statistics~\cite{gelman2014bayesian,carpenter2016stan}, generative models~\cite{goodfellow2014generative}, and probabilistic programming\slash induction~\cite{lake2015human}.

However, there has been recent criticism that many of these approaches to building more intelligent software are too far from human-level intelligence and, thus, are not likely to be enough~\cite{lake2017building,marcus2018deep}. Instead the critics argue that we actually need algorithms that build and extend causal models, can learn from very few examples (one- or few-shot learning), and can reason symbolically with the patterns and knowledge they extract from sensors~\cite{lake2017building,marcus2018deep}.

Regardless if the current set of AI technologies will be enough to reach human-level intelligence or not it is clear that software systems will increasingly incorporate them as components and sub-systems. The form of the solutions produced from these AI\slash ML technologies often look inherently different from the software that is normally developed and deployed. Thus, not only does the AI technology itself change quickly and at an increasing pace, the solutions it provides typically look very different from what software organisations and engineers are used to. This poses a new and unique set of risks and opportunities for software organisations and they need to understand and analyse these risks to select appropriate strategies.

This hybridisation of AI\slash ML and software engineering is inevitable also in another sense. There will be ample opportunity to apply AI and ML models to improve software development itself. Software engineers are close to these technologies and are likely to be early adopters in applying them on their own problems, methods and tools. This is also helped by the trend that AI\slash ML technologies is increasingly componentised and can be more easily used and reused, even by non-experts. Advances in software engineering allow AI technologies to be packaged and easily reusable through RESTful APIs\footnote{\url{https://bigml.com/api}} as automated cloud solutions, which can use multiple technologies before selecting and automatically tuning one\slash several\footnote{\url{https://cloud.google.com/automl}}. As AI becomes more accessible its use can be expected to increase even more.

Given the current expansion of the field and the large number of different ways that AI\slash ML can be applied during the engineering of software and in the software systems themselves, there is a risk of confusion and miscommunication. If we do not have shared terms to describe the terrain and an overview of possibilities it becomes harder for engineers and software organisations to properly assess risks and discuss mitigation strategies. In our experience, of working with and applying AI\slash ML technologies to software engineering problems in industry, we have seen this first-hand~\cite{feldt1998generating,feldt1999genetic,baumer2008predicting,afzalT08pred,de2013searching,feldt2013finding,feldt2014system}. 

Clear definition of terms and a way to classify and understand opportunities and risks can be critical in improving communication and enabling more detailed weighing of alternatives. This forms the basis of creating organisational strategies. Here, we propose an initial taxonomy with which to analyse and understand different ways of applying artificial intelligence and machine learning in software engineering. To show the utility of the approach we apply it to a sample of papers that have been published at the RAISE workshop on realizing synergies between AI and software engineering.

%% file: files/background.tex
\section{Background and Related Work}
\label{s:background}
Since the early days when Barr and Feigenbaum~\cite{barr1981handbook} discussed the possibilities of combining AI with software engineering, there have been some attempts to classify the field to systematically be able to attack key challenges. In 1987, Barstow~\cite{Barstow87} presented a review of how one should apply AI techniques to software engineering problems. He distinguished between programming-in-the-small (by individuals or very small groups) and programming-in-the-large (by very large groups of people). He divided AI usage into five broad categories: Software engineering methodologies, programming techniques, the architecture of the target machine, the application domain, and the history of the target software.

At RAISE'12, Clifton et al.~\cite{Clifton12RAISE} provided an overview of machine learning and software engineering in health informatics by presenting ongoing work from several projects. Many of the presented cases show that the scale of clinical practice requires new engineering approaches from both disciplines. This is but one example where one sees that both AI and SE need new engineering approaches for specific domains. However, we also see a need to address challenges on a higher level of abstraction as Harman~\cite{Harman12} pointed out in his RAISE keynote that same year. From Harman's perspective, the abstraction would serve us in developing ``strategies for finding solutions rather than the solutions themselves.'' In this particular case, we are not looking specifically at finding strategies by the use of abstraction, but rather to provide researchers and practitioners with a view of AI in SE, and how `embedded' we would wish for AI techniques to be, i.e., answering the why and what, while taking risk into consideration.

On that note, Davis et al.~\cite{Davis2016dre}, propose a taxonomy of AI approaches for adaptive distributed real-time embedded (DRE) systems. The taxonomy classifies AI approaches according to five properties needed for adaptive DRE systems: $i$) supporting a distributed environment, $ii$) supporting real-time requirements, $iii$) supporting an embedded environment, $iv$) robustly handling new data, and $v$) incorporating new data into the approach as it becomes available while the system is running. Their classification is fine-grained and is limited to the context of DRE system, since the goal is identifying suitable AI applications.

Another example of a taxonomy tailored to specific AI applications in SE is proposed by Charte et al.~\cite{Charte2018autoencoders}. Their taxonomy provides a broad view of autoencoders (AE) which are Artificial Neural Networks (ANNs) that produce codifications for input data and are trained so that their decodifications resemble the inputs as closely as possible. The AEs are used in different applications related to SE such as data compression, hashing and visualisation. The taxonomy is based on the different features of an AE, such as lower dimensionality, or noise tolerance. However, Charte et al.\ do not include AI technologies beyond AEs.  

We could not find other taxonomies that covers classification within both areas of AI and SE; existing proposals focus on one or the other and are rather fine-grained. Unlike the existing taxonomies, the one we present in this paper, AI-SEAL (Artificial Intelligence in Software Engineering Application Levels), aims to be more general and allow classification beyond a specific subject matter, such as for a specific type of software system, e.g., DRE systems, or a specific type of AI technology, e.g., AEs. It has three main facets (dimensions) to explore different perspectives of AI in SE on a more general level, thus allowing us to cover more AI approaches and be more inclusive of other application domains within software engineering. By being more coarse-grained it can also be complemented by more detailed taxonomies, such as the ones described above, or additional, lower-level facets, as outlined at the end of the next section.

AI-SEAL also differentiates itself by targeting a different use case. While the taxonomies and categorisations we have described above focus on providing an overview of the field or outlining possibilities when applying AI technologies in SE, our overall aim is to support companies and organisations developing strategies for such applications. The main aim is thus to be able to estimate and analyse the risks and costs involved, not only the possibilities. When a certain type of technology is new there is a tendency to focus on possibilities and the upside of its application while real-world adoption requires a deeper understanding also of the downsides. Our focus is on helping organisations better understand the risks.

\subsection{Methodology}
In order to propose AI-SEAL we investigated the creation and usage of taxonomies in SE\@. There are several taxonomies proposed for SE in all of its different knowledge areas (e.g., requirements and testing~\cite{Unterkalmsteiner2014taxonomy}), but very few are created systematically~\cite{Usman2017taxonomy}. Usman et al.\ performed a systematic mapping on the use of taxonomies in software engineering, and proposed a method to develop such taxonomies~\cite{Usman2017taxonomy}, which we used when defining AI-SEAL\@. 

There are four phases in the Usman et al.\ method: $i$) planning, $ii$) identification and extraction, $iii$) design and construction, and $iv$) validation. During planning we decided to be inclusive of all knowledge areas within software engineering (e.g., testing, requirements, processes) and rather seek a more fundamental aspect of how AI is applied. We argue that this is natural since it would not matter much to the risks involved whether one applies AI to for example requirements or design; we argue that the main risks arise if the actual software itself changes shape and to what degree the engineers or, even later, the users, can change the proposals and\slash or decisions put forward or implemented by the AI or software created by it.

The goal with the identification and extraction phase is to identify the main categories and associated terms used in the taxonomy. The challenge was to identify terms and categories pertaining to the different SE knowledge areas and the variety of available AI technologies, not to mention that both fields are constantly changing and evolving. Based on our own experience from applying AI in SE, and from reading relevant papers published in the last years, we identified many key facets\slash dimensions but then filtered them down.

We extracted three key facets for our taxonomy: point of application, automation level and AI technology. We decided to use a faceted analysis because this type of classification structure is suitable for new and evolving fields, since complete knowledge related to the subject area is not required or available~\cite{Usman2017taxonomy}.

The design and construction phase presents the different levels within those facets, how they were chosen and to what extent they are connected to the taxonomy's purpose and usage (we explain the details in Section~\ref{s:ai-seal}). Lastly, the validation phase aims at demonstrating how the subject matter can be classified using the proposed taxonomy. Literature provides three distinct validation methods: Orthogonality demonstration, benchmarking and utility demonstration~\cite{Usman2017taxonomy}. We chose the latter and classified a set of AI applications reported in previous editions of the Workshop on Realizing Artificial Intelligence Synergies in Software Engineering (RAISE). Additionally, utility demonstration is the most reported method to validate taxonomies in SE, and it allows us to showcase the classification extent of the taxonomy~\cite{Usman2017taxonomy}.

%% file: files/ai_seal_taxonomy.tex
\section{Taxonomy for AI-in-SE Application Levels}
\label{s:ai-seal}

\subsection{Overview and Description}
\label{ss:overview}

The purpose with the Artificial Intelligence in Software Engineering Application Levels (AI-SEAL) taxonomy is to support researchers and practitioners to communicate, understand and discuss the pros and cons of applying AI approaches when developing and in running software systems. We argue that since the end goal of any software engineering process is to deliver a running software system we cannot exclude the actual use of AI during system execution. 

An explicit goal was to keep the number of facets to a minimum; in the end we propose only three. Even though they can be further sub-divided and additional facets can help detail them more we argue that this can be left for future work. A simpler taxonomy is more likely to be useful, in particular for practitioners and companies. The three facets we propose as critical are Point of Application (PA), Type of AI (TAI) applied, and Level of Automation (LA) offered.

\begin{figure}
    \centering
    \includegraphics[width=\columnwidth]{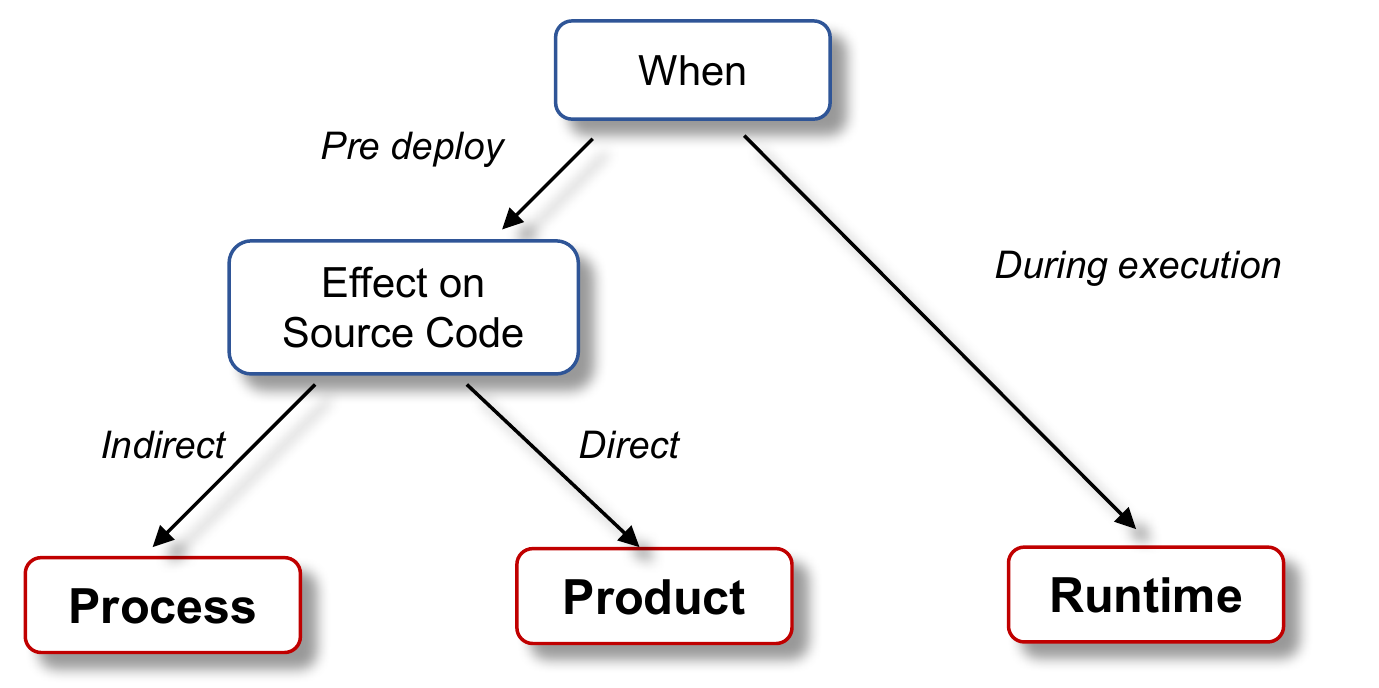}
    \caption{Overview of the different points of application (PA) in the AI-SEAL taxonomy.}
    \label{fig:ai-seal-poapp}
\end{figure}

The \textit{point of application} (PA) includes both the `when' (in time) and the `on what' (location) the AI technology is being applied (Figure~\ref{fig:ai-seal-poapp}). There are three major levels of this facet, two that are relevant before deployment of the software system (\textit{process} and \textit{product}) while the third is post-deployment representing the \textit{runtime} application of AI in a software system.

The \textit{process} level indicates that the AI is applied in the software development process and does not necessarily affect, directly, the source code that will be deployed. An example would be test analytics, which could be used to optimise testing, but it does not by itself directly alter the code, e.g.,~\cite{AfzalT08,feldt2014system}. In contrast, the \textit{product} level indicates that the AI directly affects the source code. A concrete example would be automated program repair, which manipulates the code directly to automatically fix defects~\cite{Monperrus2018survey}.

The \textit{runtime} level represents AI applications that affect the deployed software system during runtime. The canonical example would be autonomous and self-adaptive software systems in which some AI technology is learning and changing the system itself in a feedback loop~\cite{truszkowski2004nasa}. A more mundane, but recent example, would be the online learning of more optimal data structures and database indices based on the actual data stored during operation, in line with recent results from Google~\cite{kraska2017case}.

Some applications can span these main levels of PA or can be viewed as borderline between levels. An example would be autonomous driving software that includes an ANN\@. In a situation where the AI develops a part of the code, which is then compiled into the binary that goes into the final product, it would be classified under the \textit{product} application level. However, if the ANN is dynamically updating itself using runtime information from the executing software, then its PA classification would be set to the \textit{runtime} application of AI\@. For risk analysis we argue it makes more sense to then select the latter (higher) level, i.e., runtime over product and product over process. The reason is that the higher the level the less time there is, in general, for humans to intervene or even to analyse the result of the applied AI technology.

The next facet of AI-SEAL covers the \textit{Type of AI} (TAI) that is applied. Since there is not even a consensus around what AI is it becomes hard to propose a particular and stable set of levels for this facet. This facet is, thus, most likely to need to change as progress is made and new types of AI approaches are proposed. As a starting point we propose that `the five tribes of AI' classification introduced by Domingos~\cite{Domingos2015master} can be useful:

\begin{itemize}
\item Symbolist, e.g., inverse deduction. 
\item Connectionist, e.g., backpropagation.
\item Evolutionaries, e.g., genetic programming.
\item Bayesians, e.g., probabilistic inference.
\item Analogizers, e.g., kernel machines.
\end{itemize}

Even though these five tribes capture general types of AI technology it is clear that the TAI facet can be made more detailed and divided into further sub-dimensions, depending on the representations, algorithms, and artefacts used in a particular application of AI\@. However, we argue that specifying sub-dimensions hinders the practical use of the taxonomy since it can be confusing to precisely distinguish among the different existing algorithms, mainly if more than one AI technique is involved in, e.g., the product. In other words, AI-SEAL users can choose to go deeper into that facet within their domain-specific constraints, but we do not incorporate those sub-dimensions into the taxonomy itself. The main purpose of the TAI facet is to consider the particular properties of the applied AI technology and how they interact with the PA and LA facets in a particular application. For example, in a product application the risk might be much higher with using a Connectionist AI technology, which produces an opaque neural net that is hard to analyse and test, than if using an evolutionary search process to find decision rules that are short and can be analysed before they are deployed.

The last facet is \textit{the level of automation}, i.e., LA, which the AI application aims at or achieves. We base the levels of this facet on the Sheridan-Verplanck 10 levels of automation, an existing taxonomy from Automation\slash HCI research that focuses on human-computer decision making \cite{Frohm2008automation,Sheridan1980taxonomy} (Table \ref{tab:loa}).

\begin{table}
\caption{Levels of automation (LA) of decision and action selection (from \cite{Frohm2008automation} and \cite{Sheridan1980taxonomy})}
\label{tab:loa}
\begin{tabular*}{\columnwidth}{p{0.01\columnwidth}p{0.93\columnwidth}}
\hline
10. & Computer makes and implements decision if it feels it should, and informs human only if it feels this is warranted. \\
9. & Computer makes and implements decision, and informs human only if it feels this is warranted. \\
8. & Computer makes and implements decision, and informs human only if asked to. \\
7. & Computer makes and implements decision, but must inform human after the fact. \\
6. & Computer makes decision but gives human option to veto before implementation. \\
5. & Computer offers a restricted set of alternatives and suggests one, which it will implement if human approve. \\
4. & Computer offers a restricted set of alternatives and suggests one, but human still makes and implements final decision. \\
3. & Computer offers a restricted set of alternatives, and human decides which to implement. \\
2. & Computer offers a set of alternatives which human may ignore in making decision. \\
1. & Human considers alternatives, makes and implements decision.\\ 
\hline
\end{tabular*}
\end{table}

The Sheridan-Verplanck taxonomy conveys how different human operators (e.g., a developer, tester, user, or any stakeholder in the software system) and the technical system (e.g., AI technology) should cooperate by sharing the control of determining and selecting options to implement tasks. At lower levels of automation, the AI technology simply provides data through, for instance, dashboards with descriptive and visual information, while the stakeholder responsible for understanding the information and determining the next course of action of the software.

As we climb up these levels, we allow the AI technology to be more autonomous by either allowing it to suggest alternatives to the human operator (Level 2) or even, at the top level, implement the decisions itself and only inform the human if it so decides (Level 10). Hence, the higher the level of automation, the more autonomous the AI technology becomes in making decisions related to the element, e.g., product, process, or runtime, where it is being applied. And, we argue, the higher the risk involved~\footnote{We admit that there is also a case to be made that ultimately, and especially for some tasks, solutions based on AI might be less risky than having humans have the final say before taking action. However, we argue that the decision to trust AI solutions will have to be based on other risk-reduction strategies such as taking a general decision to trust certain types of systems based on empirical evidence of their safety. We thus see this as a future extension possibility rather than a threat to the taxonomy presented here.}.

\begin{figure}
    \centering
    \includegraphics[width=\columnwidth]{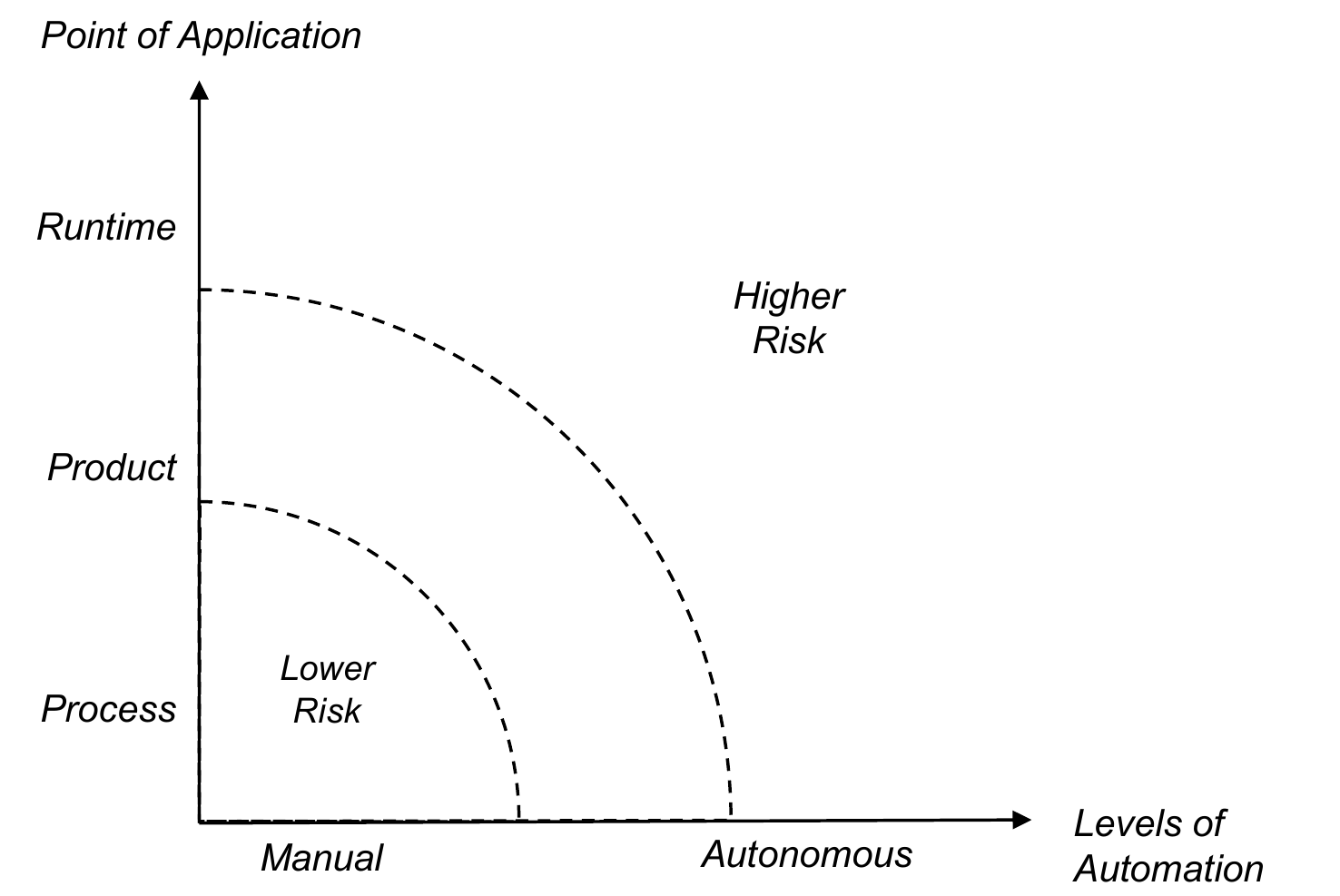}
    \caption{Levels of risk on the AI-SEAL taxonomy with respect to Points of Application (PA) and Levels of Automation (LA).}
    \label{fig:ai-seal-summary}
\end{figure}

Note that one of the main benefits of the three facets of the taxonomy is to allow practitioners and researchers to classify and understand the risks involved in the AI application. If we consider the PA facet, we argue that the risk of applying AI increases as we move from \textit{process} to the \textit{runtime} level. At each step, there is more at stake since negative consequences due to the application of inappropriate AI technologies are worse\slash costlier to reverse closer to (or after) deploying the product. As such, PA is closely related to the size of the impact as well as level of control that developers have on the AI application. Similarly, the LA facet is riskier at higher levels, since stakeholders will have less time to reverse decisions when the AI has a higher degree of autonomy (see the risk boundaries in Figure \ref{fig:ai-seal-summary}).

Thus, if an AI technology is new to a company, practitioners should start at low levels of automation (LA) to allow more human intervention, as well as at a `lower' point of application (PA), where potential issues with introducing AI will not be directly introduced to the source code. Then, by building more experience one can expand to explore higher levels of automation and points of application of the AI technology in SE and software systems.

%% file: files/evaluation.tex
\section{Evaluation}
\label{s:application}
We evaluate the AI-SEAL taxonomy through a utility demonstration~\cite{Usman2017taxonomy} by classifying papers from previous editions of the RAISE workshop. The workshop focuses on papers showcasing applications of AI in SE, in a broad sense, and is, thus, a suitable venue for identifying relevant papers.

The first instance of RAISE was held in 2012 and, except for 2017, it has been running annually (2012--2016). A total of 44 papers have been presented at RAISE over the years. Based on the title we excluded ten papers that we assessed as not presenting a specific application\slash solution of AI in SE, e.g., surveys, or papers presenting challenges or open issues. For the remaining set of papers we then included the most recent ones, published at RAISE 2015 (six out of seven papers; one not found in the IEEE database) and 2016 (four papers), as well as a random sample of another five for a total of 15 papers\footnote{\url{https://goo.gl/ERPoUk}}. We then applied the AI-SEAL taxonomy to these 15 papers. In Table~\ref{tab:results} we show an overview of the results and below we describe four papers in more detail, to show the value of our approach. 

\begin{table}
\caption{Classification of 15 RAISE papers according to the AI-SEAL taxonomy. Papers \#6 and \#15 both discuss Runtime and higher levels of automation of the AI so we mark them as borderline FW (Future Work) below and list them twice.}
\label{tab:results}
\begin{tabularx}{\columnwidth}{c c c c c}
\hline
\multirow{2}{*}{ID} & \multirow{2}{*}{Reference} & Appl.\ Point & Type of AI & Level of Auto\\
 & & (PA)    &     (TAI)  &    (LA) \\
 
\hline
\#1 & \cite{1} & Process & Analogizer & 2\\
\#2 & \cite{2} & Process & Analogizer & 4\\
\#31 & \cite{31langer} & Process & Connectionist & 7\\
\#34 & \cite{34Roychoudhury} & Process & Symbolist & 9\\
\#36 & \cite{36schindler} & Process & Analogizer & 2--3\\
\#37 & \cite{37papadopoulos} & Process & Symbolist & 2--3 \\
\#38 & \cite{38Hamza} & Process & Symbolist & 2\\
\#40 & \cite{40diamantopoulos} & Process & Evolutionary & 2--3\\
\#41 & \cite{41didar} & Process & Analogizer & 2--3\\
\#42 & \cite{42musco} & Process & Analogizer & 2--3\\
\#43 & \cite{43akbarinasaji} & Process & Analogizer & 2--3\\
\#44 & \cite{44misra} & Process & Analogizer & 2--3\\

\hline
\#6 & \cite{6}  & Product & Symbolist  & 4\\
\#15 & \cite{15} & Product & Analogizer & 7\\
\#35 & \cite{35landhaeusser} & Product & Symbolist  & 9\\

\hline
\#6 & \cite{6}  & (Runtime)\textsubscript{FW} & Symbolist  & (8)\textsubscript{FW}\\
\#15 & \cite{15} & (Runtime)\textsubscript{FW} & Analogizer & 7\\

\end{tabularx}
\end{table}

Paper \#1, written by Iliev et al.~\cite{1} uses an ontology, based on design information provided by stakeholders, to automatically predict the severity level of a defect. The AI suggests the severity levels to stakeholders who, in turn, can accept or ignore the suggestions. Therefore, the AI-SEAL classification of Paper \#1, for PA, TAI and LA is, respectively, \textit{Process}, \textit{Level 2 of Automation} and included in the \textit{Analogizer} tribe since the design information and the classification rules used by the ontology are defined in advance by stakeholders.

Paper \#6, written by de Souza Alcantara et al.~\cite{6} presents an approach where a tool learns a set of gestures that UI designers can use to design gesture-based applications for multi-touch devices. We classify this as a \textit{Symbolist} AI technology since it analyses the relations between and reasons about the different steps of a specific gesture, which are inferred based on a set of primitives. Hence, the designer first trains the tool to learn a set of gestures that can be used when designing the UI of the application. Then, during the actual design of the UI, IGT can detect gestures outside the standards and prompt the designer to ask whether the drawn gesture was a mistake. Authors state that the gestures definitions are not updated; therefore, the levels for PA and LA are, respectively, \textit{Product} (the final gesture identification output from the tool in included in the developed application) and \textit{Level 4}. However, authors discuss future work where the interaction between designer and the tool is higher, and the gestures definitions are updated; this could move the technology to higher levels of LA and, possibly, PA.

Paper \#15, written by Heitmeyer et al.~\cite{15}, is titled \textit{High assurance human-centric decision systems} and proposes an approach where AI techniques are used to detect and assist operators of a decision system that start to feel overloaded given the complexity of tasks in the decision system. They use two AI techniques, both from the \textit{Analogizers} tribe, to predict human overload based on the past interactions of the operator with the system. Ultimately, the AI application should take over the system operation while alerting the operator (\textit{LA = 7}). Therefore, the application is at the \textit{product} level, since the overload model does not update during runtime, even though authors plan to address this issue in future work. This is an example how the application can actually begin at lower risk levels, and then evolve to higher risk areas of the AI-SEAL taxonomy.

Paper \#31, written by Langer and Oswald~\cite{31langer}, is titled \textit{A self-learning approach for validation of communication in embedded systems}. The authors use neural networks (\textit{Connectionist}) to learn which communication traces are valid in a distributed embedded system. The proposed approach is used for automated integration testing focusing the communication between the distributed components, thus being applied at the \textit{process} level. Then, the AI automatically analyses the communication traces, and notify users only if an invalid communication trace is found (\textit{LA = 7}).

\begin{figure}
    \centering
    \includegraphics[width=\columnwidth]{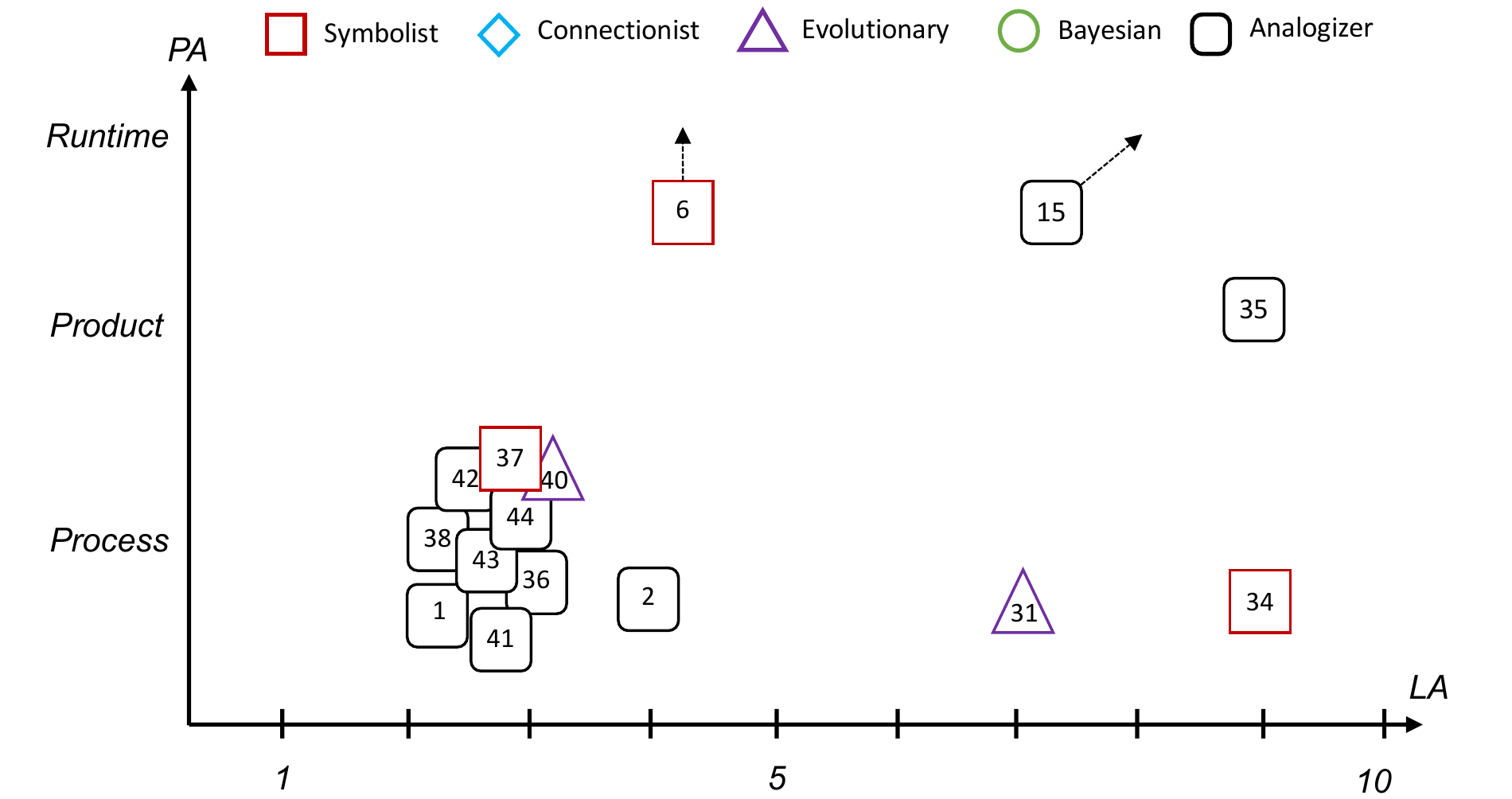}
    \caption{AI-SEAL classification of papers from different editions of RAISE\@. The papers used in this classification were \cite{1,2,31langer,34Roychoudhury,36schindler,37papadopoulos,38Hamza,40diamantopoulos,41didar,42musco,43akbarinasaji,44misra,6,15,35landhaeusser}.}
    \label{fig:papers}
\end{figure}

We summarise the classification of the papers in our taxonomy in Figure~\ref{fig:papers} and Table~\ref{tab:results}. Most of the papers ($53\%$) use analogizers, and, regarding the PA facet, most of them are applied at the process level ($80\%$), and at lower levels of automation. In contrast, none of the evaluated papers proposed applications of AI technologies from the Bayesian tribe, and only two of them discussed Runtime application of the AI as future work. We argue that this more might show a bias of the RAISE workshop than anything fundamental; for example we gave two papers as examples of Runtime applications when introducing the facet. An explanation for the little use of Bayesian solutions might be that they are not as clearly seen as part of AI technology as part of advanced statistics; however, as Domingos makes clear in his book their view of AI is also valid and general~\cite{Domingos2015master}.

We note that this is an initial proposal on how the taxonomy can be used to raise awareness of current state of art regarding AI applications in SE\@. Certainly, some applications will be more challenging than others to classify, particularly when the application or technology span across different levels of one or more facets or as new AI technologies and solutions are introduced. 

\section{Discussion}
\label{s:discussion}

We have introduced a taxonomy (AI-SEAL) having three facets that can be used to classify and analyse ways of applying AI in software engineering and thus help understand the associated risks and opportunities. Our main argument is that the risk one takes in applying AI relates to the level of control and the time given to exert control that the developers and users have (over) the decisions proposed or taken by the AI component. Together this explains two of the three facets; the third is used for basic characterisation of the AI technology being applied.

Our evaluation shows that AI-SEAL can classify applications regardless of the SE knowledge area (e.g., design, development, maintenance) and even domain-specific information (e.g., UI design, safety-critical systems, or object-oriented programming) involved. This is by design; we argue that a taxonomy needs to have this property to be generally useful. It can of course be complemented by additional facets or existing taxonomies to further detail the two main facets.


Similarly, the PA facet is inclusive of different application domains and types in SE, since it depends on elements present in any SE application, i.e., the source code of the system, that is eventually deployed, and a process to develop it. There are plenty of ways to expand on the PA facet, for example by using SWEBOK's knowledge areas to further describe what the AI is applied to and when. In fact, this possibility to update and expand the facets independently is one of the main benefits of a faceted taxonomy~\cite{Usman2017taxonomy}, thus being a suitable design choice for a taxonomy in dynamic fields such as AI and SE\@.

The demonstration (Figure~\ref{fig:papers}) also reflects the risks of the classified applications. For instance, Paper \#15 is an example of a software system controlled by an AI and a human operator that cooperate to control unmanned air vehicles (UAV), an example of autonomous vehicle, which are widely used by the military for surveillance and targeting~\cite{15}. Therefore, the risk related to that AI in SE application is high since mistakes caused by the AI could have adverse outcomes.

On the other hand, Paper \#1 \cite{1} has lower risk, where the AI automatically classifies the level of severity of found defects. Note that at $LA = 2$ the AI provides input, but stakeholders are the ones responsible to decide, so the development process can recover from eventual adversities with poor classification. Certainly, at higher levels of automation (e.g., LA $\geq$ 6), the risks would be higher since the AI becomes more autonomous and can disrupt the development process or lead to decisions that ultimately decrease quality.

On that note, the risks in Papers \#2 and \#6 differ mainly due to the PA levels. Paper \#2 detects clones on source code during maintenance, thus having an impact on maintenance costs and internal quality (e.g., refactoring) as opposed to paper \#6 that is used to assist its end-user (UI designers). Any mistakes in the AI application in Paper \#6 affects the product, thus its impact on external quality is higher.

Certainly, the risk analysis is sensitive to other factors involving both the AI and SE parts. For example, the familiarity that the company and the engineers have with the particular AI technology being used will be important. However, the value of AI-SEAL is to give an overview of the risk and the possible impacts\footnote{We also see a potential to analyse the overall `reach' of the decisions/proposals of AI component in the wider system in a more detailed risk analysis. Since the ultimate risk might not be high, despite a high LA value, if the impact of the decisions are limited by the rest of the system.}. Therefore, we recommend practitioners interested in exploring possible AI application in their SE projects to begin in the lower risk areas, and then move towards riskier areas as they gather experience on the AI application in SE\@. For researchers, AI-SEAL is useful to map the field and identify areas to employ more research effort.

In it's current form AI-SEAL does not help in specific classification of AI technology, for example helping practitioners to understand the distinction between different ANN models, or exposing the trade-offs when applying machine learning algorithms on different knowledge areas in SE\@. Instead, our taxonomy aims to support practitioners and researchers in understanding the high-level aspects and the impact of their AI applications. That decision is more on the strategic level rather than a solution instance (e.g., which ANN to choose from).

In general we agree with Harman that compartmentalising and deconstructing AI for SE into sub-domains is tempting but would be a mistake~\cite{Harman12}. The SE community can benefit significantly by discussing the strategies rather than the solutions themselves, and we believe that AI-SEAL can help in this endeavour.

%% file: files/conclusions.tex
\section{Concluding Remarks}
\label{s:conclusions}

In this paper we propose the AI-SEAL taxonomy to help researchers and practitioners to classify different AI applications in software engineering. The taxonomy has three facets allowing its users to classify, the point of application (process, product and runtime), the type of AI technology (based, initially, on the five tribes proposed by Domingos~\cite{Domingos2015master}) and the level of automation of the applied technology (inspired by Sheridan-Verplanck's 10 levels of automation~\cite{Sheridan1980taxonomy}).

Besides helping its user to understand the field of AI applications in SE, AI-SEAL provides a basis for software engineers to consider the risks of applying AI\@. This advantage allows, for instance, practitioners to reason about the trade-offs of introducing the AI technology in their processes and products. In addition, the taxonomy is not constrained by domain-specific applications, thus covering all knowledge areas of software engineering. 

We demonstrate the use of the taxonomy by classifying 15 papers from past RAISE workshops. Most of the papers focused on supporting stakeholders during the development process but did not directly affect the source code or the runtime behaviour of the systems. There was also an uneven use of the many different AI approaches that exist; in particular a lack of Bayesian and, surprisingly, Connectionist (neural net) ones.
Future work includes classification of more papers as well as proposing more detailed and explicit support for risk analysis.